\documentclass{pasj02} 
\usepackage{amsmath}
\usepackage{url}
\jyear{2025}
\Received{}
\Accepted{}
\Published{}

\begin{document} 
\title{X-ray Doppler tomography of Fe~K$\alpha$ emission in a low-mass X-ray binary
4U~1822--371 --- a localized reflector at the accretion stream--disk overflow}
\author{Naoto~\textsc{Sameshima}
\altaffilmark{1,2}\altemailmark
\orcid{0009-0005-6772-5529}
\email{naoto@ac.jaxa.jp}
}
\author{Masahiro~\textsc{Tsujimoto}
\altaffilmark{1,2}\altemailmark
\orcid{0000-0002-9184-5556} 
\email{tsujimoto.masahiro@jaxa.jp} 
}
\author{Makoto~\textsc{Uemura}
\altaffilmark{3}
\orcid{0000-0002-7375-7405}
}

\altaffiltext{1}{Institute of Space and Astronautical Science, Japan Aerospace Exploration Agency, Chuo-ku, Sagamihara, Kanagawa 252-5210, Japan}
\altaffiltext{2}{Department of Astronomy, Graduate School of Science, The University of Tokyo, Bunkyo-ku, Tokyo 113-0033, Japan}
\altaffiltext{3}{Hiroshima Astrophysical Science Center, Hiroshima University, Higashi-Hiroshima, Hiroshima 739-8526, Japan}
\KeyWords{stars: neutron --- techniques: radial velocities --- X-rays: binaries --- X-rays: individual (4U~1822--371)}  

\maketitle

\begin{abstract}
 We present the X-ray Doppler tomography of the Fe~K$\alpha$ (6.4 keV) fluorescence line
 of the low-mass X-ray binary 4U~1822--371 obtained with XRISM. Eleven orbits of this
 short period (5.57~hr) binary were covered. The Doppler shift of the line shows clear
 modulation with the orbital period, motivating us to apply the Doppler tomography in the
 X-ray band for the first time. The resulting velocity map reveals a compact feature at
 ($v_{\mathrm{x}}$, $v_{\mathrm{y}}$) $\sim$ (--550, $+$125) km~s$^{-1}$. This is
 inconsistent with the emission originating from a symmetric accretion disk, an extended
 corona around the neutron star, or the surface of the neutron or companion
 star. Instead, it suggests that the emission originates from the accretion stream-disk
 overflow. Remarkably, the Fe~K$\alpha$ velocity map closely resembles that of the O VI
 3811~\AA{}, indicating that both X-ray and optical lines arise from the same site
 irradiated by the central X-ray source. These results provide the first
 velocity-resolved X-ray map of the fluorescent line, directly localizing the major
 reflector in an X-ray binary and establishing X-ray Doppler tomography as a new probe of
 the structures of accreting systems.
 %
\end{abstract}


\section{Introduction}\label{s1}
Compact objects in binary systems ---black holes (BHs) and neutron stars (NSs)---
accrete matter from their companion star, often forming an accretion disk. A fraction of
the liberated gravitational energy is released in the form of strong X-ray radiation,
which, in turn, contributes to expelling the surrounding matter away from the
system. This is a pumping mechanism of matter and energy in the Universe that is
observed on various scales. However, it is impossible to spatially resolve such
accreting and outflowing structures in a system using direct imaging methods.

The Fe~K$\alpha$ X-ray fluorescence line at 6.4~keV can be a powerful probe of the
spatial structures of accreting systems
\citep{basko1978,makishima1986,george1991,reynolds2003}. The neutral or low-ionized Fe
in the system is photoionized by the intense X-ray emission from the compact object. The
inner-shell photoionization leaves a vacancy in the K shell, and a fraction of it decays
radiatively to emit a fluorescent line at 6.4~keV. Due to the high cosmic abundance of
Fe, the large fluorescence yield, and the sufficiently hard energy to penetrate through
an interstellar hydrogen column of $\approx 10^{22}$~cm$^{-2}$, the line is seen in many
X-ray binaries \citep{torrejon2010,tzanavaris2018} as well as active galactic nuclei
\citep{nandra2006}. However, the site of line production in accreting systems is the
subject of intensive debate, despite numerous X-ray observations made over many decades.

In the optical band, indirect imaging methods have been developed to investigate the
structures of accreting systems, in particular of binaries consisting of a white dwarf
(WD) and a late-type companion called cataclysmic variables (CVs). One of the most
successful methods is Doppler tomography \citep{marsh1988,marsh2001,echevarria2012}. The
spectral profile of an emission line (H$\alpha$ is a popular choice) changes as the
viewing angle changes along the orbit. The profile in each phase is the integration of
the projected velocity multiplied by the line emissivity across the system. Doppler
tomography is a technique that converts the trail spectrogram (a series of line profiles
as a function of phase) into a velocity map (emission distribution in the velocity
space) by applying the method of medical tomography. Different structures in accreting
systems have different velocities: some follow rigid rotation around the centre of mass
of the binary, some follow Keplerian rotation around the compact object, and some follow
the ballistic equation of motion from the first Lagrangian point (L1). They are
visualized as different distributions on the velocity map, which helps to identify the
origin of the targeted emission line.

The application of this powerful technique to X-ray lines has long been awaited
\citep{harlaftis2001}. The low-mass X-ray binaries (LMXBs), in which the WD in CVs is
replaced by an NS or BH, are natural targets. However, no X-ray spectrometers have been
available to produce trail spectrograms with the required accuracy and cadence. A
premise of Doppler tomography is that the target line needs to be spectrally
resolved. The typical velocity in accreting systems is a few hundred km~s$^{-1}$,
requiring X-ray spectrometers with a resolution of $R \gtrsim 1000$. A large effective
area ($A_{\mathrm{eff}}$) is also needed for X-ray telescopes to accumulate well-exposed
spectra at each orbital phase.

The X-ray Imaging and Spectroscopy Mission (XRISM; \cite{tashiro2024}) hosts an X-ray
microcalorimeter spectrometer \citep{kelley2025,ishisaki2025}. The spectroscopic
performance of the instrument in the Fe~K$\alpha$ band is significantly improved ($R
\sim 1400$, $A_{\mathrm{eff}} \sim 174$~cm$^{2}$) compared to the previously best
high-energy transmission grating (HETG) spectrometer \citep{canizares2000} onboard the
Chandra X-ray observatory ($R \sim 167$, $A_{\mathrm{eff}} \sim 20$~cm$^{2}$). The time
is ripe to apply Doppler tomography in X-rays, which we present for the first time in
this paper. Based on the technique, we investigate the origin of the Fe~K$\alpha$
fluorescence line in LMXBs.

\medskip

The plan of the remaining part of this paper is as follows. We describe the target in
\S~\ref{s2} and the data in \S~\ref{s3} for the instrument (\S~\ref{s3-1}) and the
observation (\S~\ref{s3-2}). We present the data analysis in \S~\ref{s4} for the light
curve (\S~\ref{s4-1}) and spectra (\S~\ref{s4-2}) and obtain the radial velocity curve
of the Fe K$\alpha$ line (\S~\ref{s4-3}). In \S~\ref{s5}, we show the result of the
Doppler tomography for the line. In \S~\ref{s6}, we compare between the X-ray and
optical results. We conclude the paper in \S~\ref{s7}.

\section{Target}\label{s2}
4U\,1822--371 (hereafter 4U1822) is a LMXB consisting of a NS with a mass $1.97 \pm 0.36
M_{\odot}$ and a late-type dwarf companion with a mass $0.50 \pm 0.06 M_{\odot}$
\citep{munoz-darias2005}. It has some unique properties suitable for the application of
Doppler tomography. First, the binary period is short (5.57~hr; \cite{mason1980}),
allowing us to cover many orbital cycles in a modest telescope time. Second, the
inclination angle is almost edge-on ($i=82.5 \pm 1.5$ deg; \cite{heinz2001}), maximising
the line-of-sight projection of the velocities in the orbital plane. Because of the high
inclination angle, the direct X-ray emission from the NS is always blocked presumably by
the outer rim of the disk, thus the reprocessed emission (e.g.; the Fe~K$\alpha$
fluorescence and electron-scattered continuum) is visible without contamination of
otherwise overwhelming direct emission full of absorption features. A handful of
accreting systems are in such a fortuitous configuration, which are called accretion
disk corona (ADC) sources \citep{white1982}. Third, 4U1822 is the only ADC source in
which the NS exhibits a coherent pulse with a period of 0.59~s. The pulse arrival times
are measured to be modulated by binary motion, leading to the determination of orbital
parameters with high precision, including an eccentricity of $<$0.031
\citep{jonker2001}. The modulation has no phase shift with respect to sharp X-ray
eclipses. Thus, the uncertainty of the phase origin defined at the superior conjunction
is negligible, which would otherwise introduce systematic errors in Doppler tomography.

4U1822 is also one of the brightest LMXBs in the optical band
\citep{griffiths1978}. Together with the ideal properties above, it is a popular target
to apply Doppler tomography using a variety of optical lines
\citep{casares2003,peris2012,somero2012}. Different lines exhibit different
distributions on the velocity map, tracing different structures in the system. The Bowen
fluorescence emission line is from the companion surface illuminated by the X-ray
emission of the NS, the He~I 4471~\AA{} line is from the first Lagrangian point (L1) but
with a slightly leading phase, and He II 4686~\AA{} is either from the accretion disk
\citep{casares2003} or the overflow from the hot spot \citep{somero2012}. Here, the hot
spot is created at the point where the accretion stream from L1 lands on the accretion
disk at its outermost radius, and a significant fraction of the matter in the stream
is bounced to form a spray-like structure \citep{armitage1998,kunze2001}, which we
hereafter call the stream-disk overflow.

4U1822 exhibits a larger equivalent width of the Fe K$\alpha$ line than other X-ray
binaries \citep{torrejon2010} and was studied extensively using low-
\citep{parmar2000,heinz2001,iaria2001,somero2012,sasano2014,niu2016,anitra2021} or
high-resolution \citep{cottam2001,ji2011,iaria2013} X-ray spectrometers. Three HETG
observations were conducted totaling 195~ks, in which many discrete narrow emission
lines of both highly-ionized ions as well as the Fe K$\alpha$ line were identified.
\citep{cottam2001,ji2011,iaria2013}. The intensity of these lines was found modulated by
the orbital motion. Velocity shifts of these lines were more difficult to constrain,
which were only poorly constrained with an error bar of $\gtrsim$200~km~s$^{-1}$ for
ionized lines or not detected for the Fe K$\alpha$ line. Based on the non-detection of
the orbital modulation of the velocity shift and broadening of the Fe K$\alpha$ line,
they argued that the line originates from an extended region in the system.

\section{Data}\label{s3}
\subsection{Instrument}\label{s3-1}
XRISM is the X-ray observatory launched in September 6, 2023 under an international
collaboration between JAXA and NASA with ESA participation. The \textit{Resolve}
instrument \citep{kelley2025,ishisaki2025} onboard XRISM \citep{tashiro2024} has an
array of 6$\times$6 X-ray microcalorimeter pixels (0--35) placed at the focal plane of
the grazing-incident X-ray mirror \citep{hayashi2024}. One of the pixels (12) is
displaced from the array and is constantly illuminated by an $^{55}$Fe source for
calibration purposes. Each microcalorimeter consists of an HgTe X-ray absorber and an
ion-implanted Si thermistor \citep{porter2025}. They are thermally attached to a 50~mK
stage, which is controlled by the two-stage adiabatic demagnetization refrigerator (ADR;
\cite{shirron2025}). They are stored in a cryostat \citep{yoshida2024}. At the top of
the cryostat along the X-ray optical path, an X-ray transmissive window made of Be is
installed. This device is to keep the cryostat leak-tight before the launch and was
intended to be opened in orbit, but has not been successful yet. Because of the window,
the energy range is restricted to above $\sim$1.7~keV \citep{midooka2021}.

\textit{Resolve} works in the photon-counting mode independently for each pixel. The
time series data of each X-ray event are correlated with a template in the time domain
to derive the energy \citep{Boyce1999}. The highest accuracy of the energy determination
is obtained for events isolated in time from others in the same pixel. They are called
the Hp grade \citep{ishisaki2018b}, which we exclusively used in this paper. An
anti-coincidence detector is placed beneath the microcalorimeter array, which is used to
discriminate events induced by cosmic rays. The residual non X-ray background is very
low with only $\lesssim$0.5 events per spectral bin (5~eV) per 100~ks exposure
\citep{kilbourne2018}, which we ignored in this paper.

The changing environment in the orbit makes the energy gain of \textit{Resolve}
variable. To track and monitor the energy gain of each pixel throughout observations,
another set of $^{55}$Fe sources is installed in the filter wheel (FW;
\cite{shipman2025}) above the cryostat. The calibration sources are rotated into the
position to illuminate the entire array several times in a typical observation. The Mn
K$\alpha$ line at 5.9~keV after the electron capture decay of $^{55}$Fe is used to
determine the energy gain at each illumination epoch, based on which the gain variation
as a function of time is modelled and applied to calibrate the energy for each event
\citep{porter2025}. Such intermittent calibration measurements during observations is
unseen in other high-resolution X-ray spectrometers and is a key to providing an
unprecedented energy resolution of $\sim$4.5~eV (FWHM) and energy determination accuracy
of $\lesssim$0.2~eV (34~ppm) at 5.9~keV \citep{porter2025,eckart2025}, which are needed
for Doppler tomography.

\subsection{Observation and data reduction}\label{s3-2}
\begin{figure*}[!hbtp]
 \begin{center}
  \includegraphics[width=1.0\textwidth,clip]{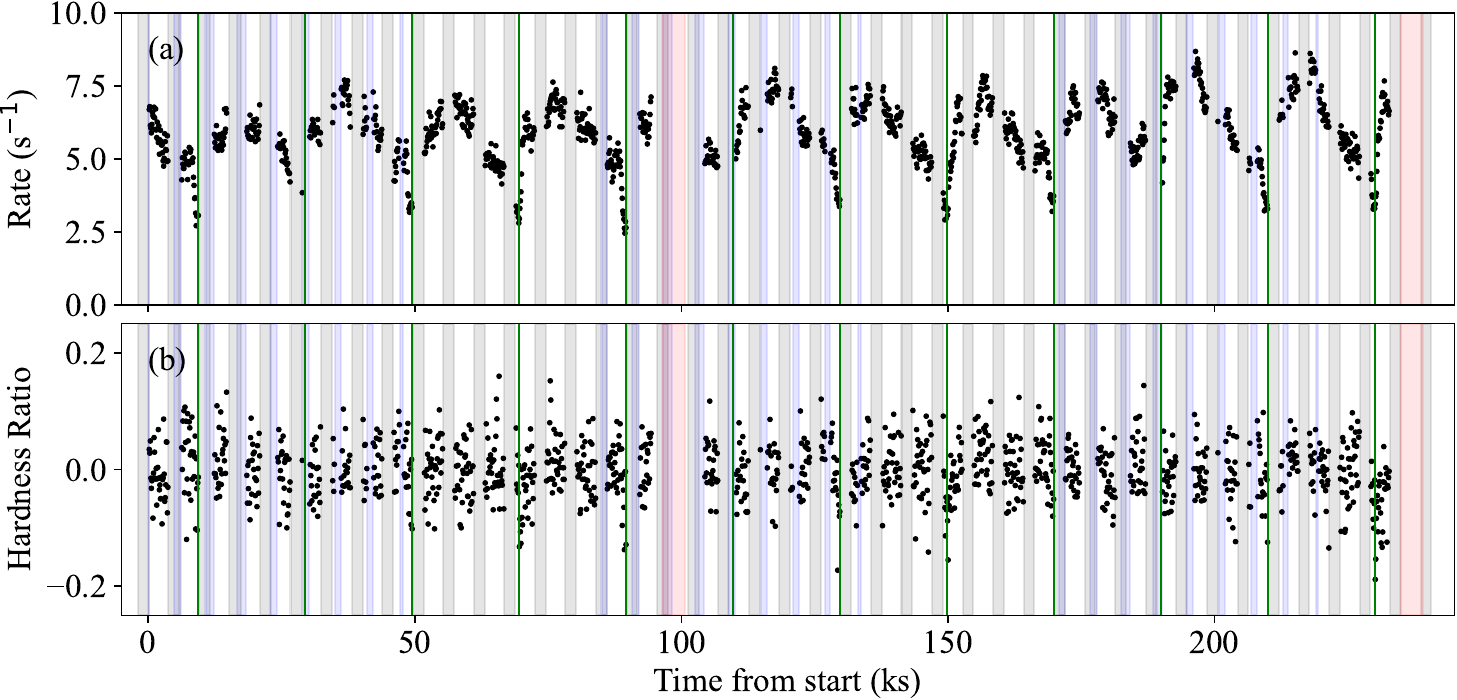}
 \end{center}
 \caption{X-ray light curves: (a) count rate (2--12~keV) and (b) hardness ratio
 (2--5.857 keV for the soft and 5.857--12~keV for the hard band). The origin of time is
 2025-04-12 18:00:00 UT (60777.75 in MJD). The observation was interrupted by ADR
 recycles (red), SAA passages (blue), and Earth occultations (grey). The eclipses at the
 superior conjunction ($\phi_{\mathrm{orb}}=0$) are shown with green lines.
{Alt text: two scatter plots of X-ray lights for the Resolve count rate and hardness ratio.}
}
 \label{f05}
\end{figure*}

4U1822 was observed with XRISM on 2025 April 12--15 (60777.8--60780.4 in modified Julian
date) for a total telescope time of 233~ks (sequence number: 201000010), equivalent to
11.6 orbital cycles (figure~\ref{f05}). The \textit{Resolve} observation was interrupted
by occultation by the Earth, the passages of the South Atlantic Anomaly (SAA) and the
ADR recycle operations for 46\% of the telescope time, yielding the total on-source time
of 125~ks.

The ADR recycling was performed twice in the middle and at the end of the
observations. A total of nine fiducial points were obtained for the energy gain
calibration during the Earth occultations, and the gain variation was successfully
modelled for all pixels. Using the Mn K$\alpha$ line, the energy resolution of the Hp
grade events was evaluated as 4.60 $\pm$ 0.02~eV (FWHM) and the systematics in the
energy determination as 0.1~eV (5~km~s$^{-1}$) as can be found in a data quality
report\footnote{The report is available at
\url{https://heasarc.gsfc.nasa.gov/FTP/xrism/postlaunch/gainreports/2/201000010_resolve_energy_scale_report.pdf}.}.

We retrieved the pipeline products processed with the processing version 03.00.013.010
\citep{doyle2022}. We used the \texttt{HEASoft} package version 6.36 for the
analysis. Starting from the cleaned event list, we removed pixel 27, which is known to
behave anomalously sometimes. We selected Hp grade events in the 2--12~keV band. We
also applied an additional event screening based on the relation between the rise time
and height of the pulse shape among X-ray events \citep{mochizuki2025}. As a result, we
obtained $7.3 \times 10^{5}$ events. The count rate is 6.7~s$^{-1}$ and the Hp branching
ratio is 0.90 on average. The response files of the detector and the telescope were
generated using standard tools and used for the spectral fitting.

\section{Analysis}\label{s4}
\subsection{Light curve}\label{s4-1}
\begin{figure}[!hbtp]
 \begin{center}
  \includegraphics[width=1.0\columnwidth,clip]{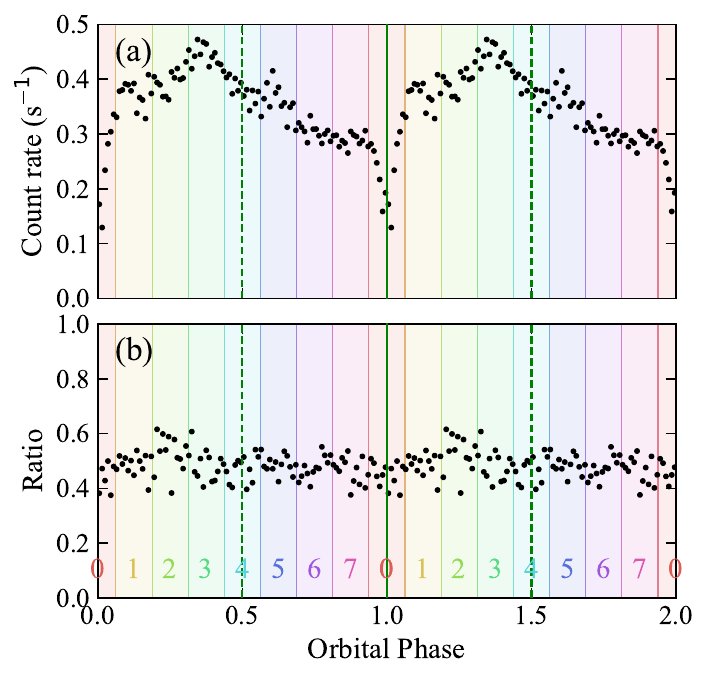}
 \end{center}
 \caption{Folded light curve: (a) count rate (2--12~keV) and (b) count rate
 ratio between the Fe K$\alpha$ line band (6.35--6.45~keV) and the neighboring continuum
 band (6.0--6.3~keV) for two orbital cycles. The superior ($\phi_{\mathrm{orb}}=0$) and
 inferior ($\phi_{\mathrm{orb}}=0.5$) conjunctions are shown with the solid and dashed
 green lines, respectively. The eight bins for time-resolved spectroscopy are shown in
 different colors.
 {Alt text: two scatter plots of X-ray lights folded by the orbital period for the
 Resolve count rate and the Fe K$\alpha$ line against continuum ratio.}
 }
 \label{f06}
\end{figure}

Figure~\ref{f05} shows the light curves of the total count (2--12~keV) and the hardness
ratio HR $\equiv (H+S)/(H-S)$. Here, $H$ and $S$ are the count rate in the soft and hard
band, respectively, with the boundary at the medium energy (5.857~keV). In the light
curve, the repeated pattern caused by orbital motion is seen, but the color hardly
changed suggesting that the spectral shape did not change much. The light curve was
folded by the orbital period ($P_{\mathrm{orb}}$) in figure~\ref{f06}. Despite the
change in the count rate by $\sim$2.5 times, the Fe K$\alpha$ line band (6.35--6.45~keV)
counts relative to the neighboring continuum band (6.0--6.3~keV) counts remained nearly
constant, showing that the line was eclipsed to the same fraction with the continuum
emission. We divided the orbital phase equally into eight bins ($i \in \{0...7\}$ for
$\phi_{\mathrm{orb}} = [\frac{i-0.5}{8}, \frac{i+0.5}{8}]$).

X-ray eclipse times of 4U1822 have been recorded for four decades in 1977--2017, from
which the ephemeris was derived \citep{mazzola2019}. We use their equation (1). The
accuracy is sufficiently high, so that the accumulated error is expected to be less than
$10^{-2}$ for the phase origin when extrapolated to the XRISM observation date. Indeed,
as shown in figure~\ref{f06}, the eclipse timing matches very well with the XRISM data
with no need for further correction.

\subsection{Spectrum}\label{s4-2}
\begin{figure}[!hbtp]
 \begin{center}
  \includegraphics[width=1.0\columnwidth,clip]{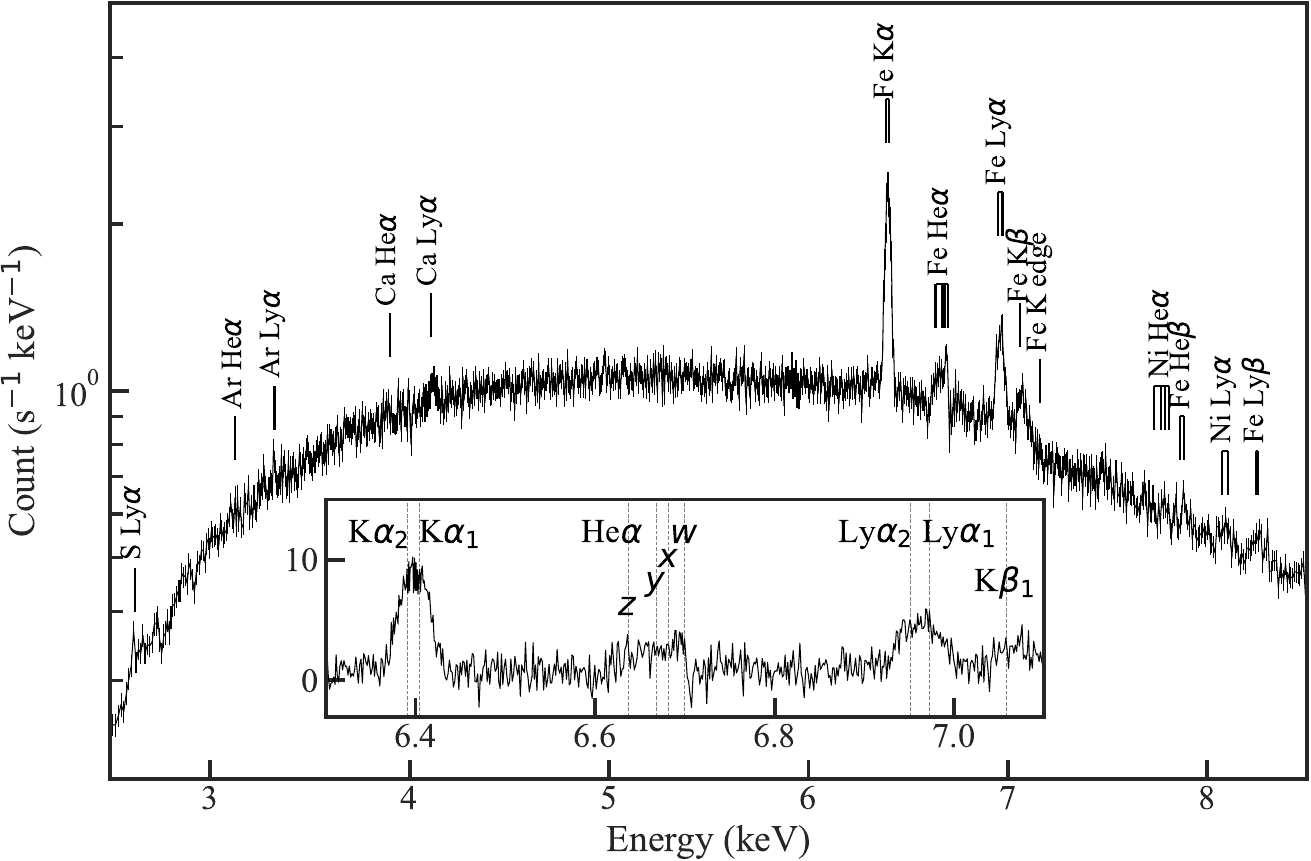}
 \end{center}
 \caption{X-ray spectrum with a coarse binning. The RV shift was corrected for
 individual X-ray photons based on their arrival times following the RV curve in
 figure~\ref{f02b}. Conspicuous spectral features are labeled. In the inset,
 continuum-subtracted spectrum of the Fe K band is shown with the dashed lines at the
 rest-frame energy.
 {Alt text: X-ray spectrum.}
 }
 \label{f07}
\end{figure}

Figure~\ref{f07} shows the X-ray spectrum. Upon the hard continuum emission, a handful
of conspicuous spectral features are found as labelled. A close-up of the Fe K-band
spectrum is shown in the inset, which contains rich features --- the Fe K$\alpha$
(6.4~keV) and K$\beta$ (7.1~keV) fluorescence as well as the Fe~XXV He$\alpha$ (6.7~keV)
and Fe~XXVI Ly$\alpha$ (7.0~keV) line complexes from highly-ionized Fe.

The Fe~K$\alpha$ line is the strongest feature. It is broad and the two major lines
K$\alpha_1$ and K$\alpha_2$ are not clearly resolved. These would have been easily
resolved with the \textit{Resolve} resolution if they were not broadened
\citep{tsujimoto2025}. The accompanying K$\beta$ line was detected without being
contaminated by the K edge feature at 7.12~keV. It is shifted blue-ward, unlike the
K$\alpha$ line. We will exploit this differential shift later.

For the features from highly ionized Fe, the Fe~XXVI Ly$\alpha$ line is also broadened.
The fine-structure doublet Ly$\alpha_1$ and Ly$\alpha_2$, which would have been easily
resolved, are not resolved. On the other hand, the Fe~XXV He$\alpha$ line is not as
broadened. Its resonance line ($w$) is resolved into emission and absorption features
(inset in figure~\ref{f07}) reminiscent of the P Cygni profile, which would be the first
example in ADC sources of LMXBs. This would support the idea that the outgoing wind
exists in this system as predicted for another ADC source of LMXBs \citep{tomaru2023}
and as observed in an ADC source of CVs \citep{bearda2002}.

\begin{figure}[!hbtp]
 \begin{center}
 \includegraphics[width=0.8\columnwidth,clip]{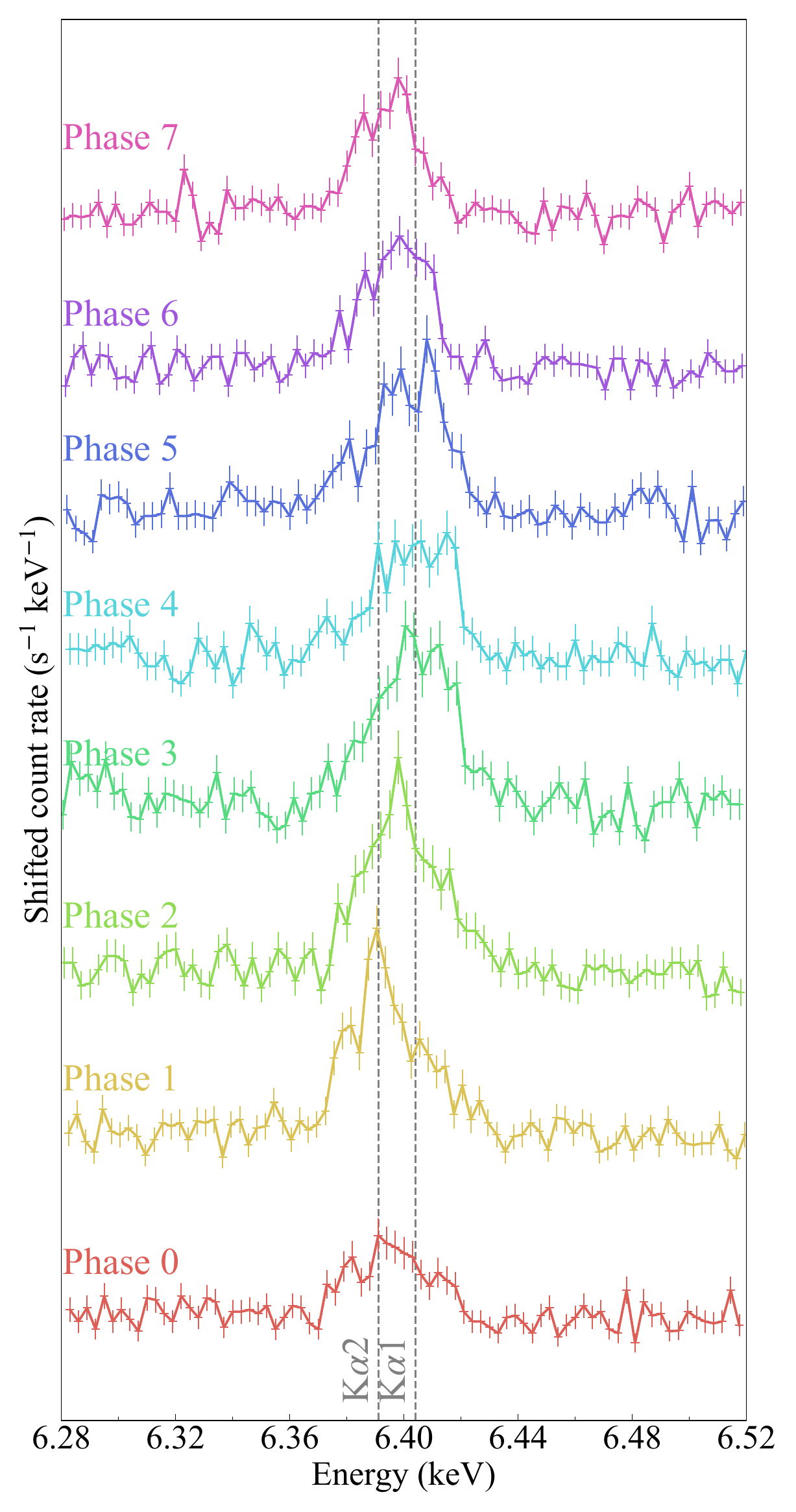}
 \end{center}
 \caption{Fe~K$\alpha$ line at eight orbital phase bins. The data (colored) as well as
 the Fe~I K$\alpha_1$ and K$\alpha_2$ line energies (dashed grey lines) are shown.
 {Alt text: Fe~K$\alpha$ line at eight orbital phase bins.}
 }
 \label{f02a}
\end{figure}

\subsection{RV curve}\label{s4-3}
\begin{figure}[!hbtp]
 \begin{center}
 \includegraphics[width=1.00\columnwidth,clip]{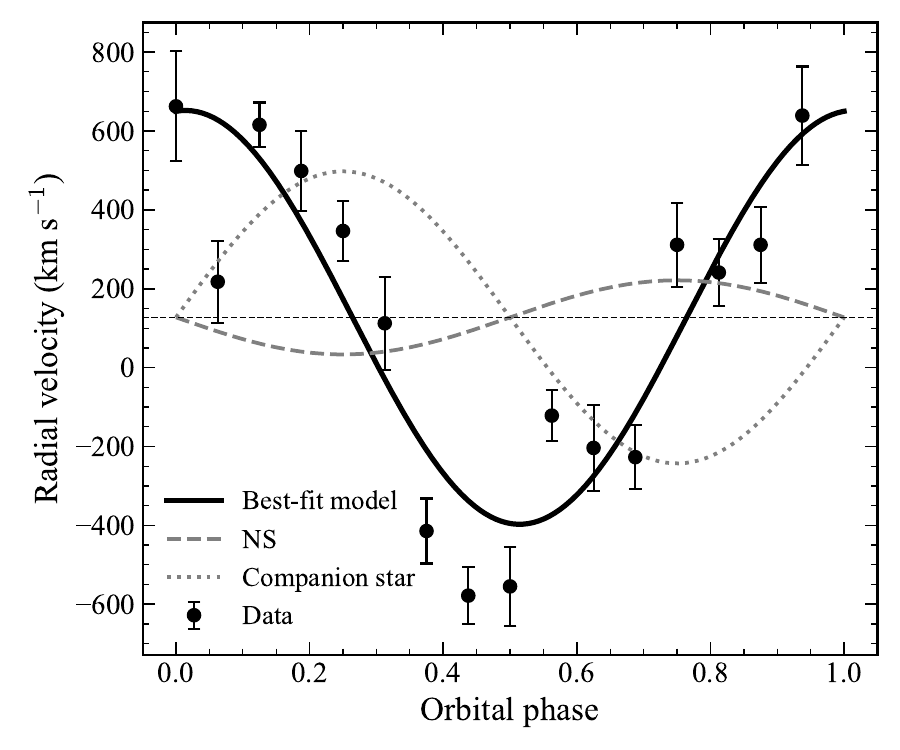}
 \end{center}
 \caption{Radial velocity curve. The observed curve of the Fe~K$\alpha$ line (black
 points for the data and black solid curve for the best-fit model), and those expected
 if the line was associated with the NS (grey dashed curve) or the companion star (grey
 dotted curve). In some phases, the data points are offset from the expected RV curve
 due to local noise in the spectrum. This is greatly alleviated by the use of Doppler
 tomography, in which the trail spectrogram is averaged along the orbital period.
 {Alt text: Radial velocity curve of the Fe~K$\alpha$.}
 }
 \label{f02b}
\end{figure}

We constructed the Fe~K$\alpha$ spectra for eight orbital phase bins (0 to 7) of equal
length (figure~\ref{f02a}). Eclipses are in the middle of phase 0
(figure~\ref{f06}). Line energy is clearly modulated, which was not detected before
\citep{cottam2001,ji2011,iaria2013}.

We constructed the radial velocity (RV) curve by tracking the peak of the spectrum,
which shows a sinusoidal pattern (solid curve in figure~\ref{f02b}). The amplitude and
phase of the RV curve are entirely different from those expected if the line was
associated with the NS (dashed curve in figure~\ref{f02b}); if it were, the largest
amplitude of 94~km~s$^{-1}$ should have been observed at phase 0.25, while it was
actually observed as 525~km~s$^{-1}$ at phase 0.51. The observed RV curve is
also different if the line was associated with the companion (dotted curve in
figure~\ref{f02b}). The data were fitted with a sinusoidal curve of a fixed period. The
best-fit parameters are: $525 \pm 32$~km~s$^{-1}$ for the amplitude, $127 \pm
23$~km~s$^{-1}$ for the offset, and --$0.26 \pm 0.01$ for the phase shift with respect
to the RV curve of the NS.


\section{Results}\label{s5}
We now apply Doppler tomography to the Fe~K$\alpha$ line to make the best use of its
line profile, not just the peak as in the RV curve. Doppler tomography is a problem to
construct the optimum velocity map, which is a collection of the intensity in each $N
\times N$ cell in the ($v_{\mathrm{x}}$, $v_{\mathrm{y}}$) space fixed to the binary
rotating frame, from the data. The optimum solution of this inverse problem is obtained
by minimising the likelihood (the distance between the data and the model) plus the
regularization term. The regularization term is intended to suppress
overfitting. Its weight relative to the likelihood term is controlled by a hyper-parameter
$\lambda$.

We used an implementation based on the total variation minimization approach --- the
\texttt{DTTVM} code version 1.11 \citep{uemura2015}\footnote{Available at
\url{https://home.hiroshima-u.ac.jp/uemuram/dttvm/}.}, which comes with optimization,
visualization, and cross validation capabilities. The regularization term is given as
the sum of intensity differences between adjacent cells in the velocity map, thereby
tending to leave sharp-edged features that may be intrinsically seen in accreting
systems. The code was developed and verified for optical data, but it worked for the
present X-ray data without issue. The only tweak we made is the line spread function, in
which we assumed that a monotonic line is split into two lines (K$\alpha_1$ and
K$\alpha_2$) with a fixed relative energy difference, natural width, and intensity. The
two lines are convolved with the instrumental line spread function with a Gaussian width
of 105~km~s$^{-1}$. The spectrum at each phase was normalized to the continuum
level. The velocity offset is taken from the best-fit value of the RV curve
(figure~\ref{f02b}).

\begin{figure}[!hbtp]
 \begin{center}
  \includegraphics[width=1.0\columnwidth,clip]{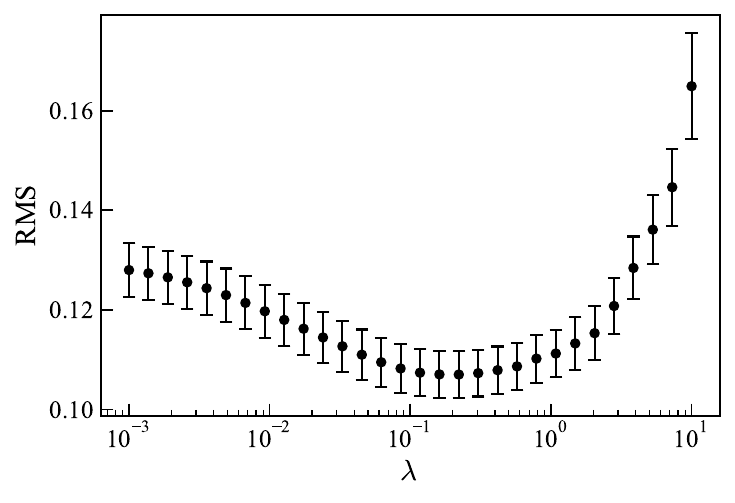}
 \end{center}
 \caption{Cross validation of the velocity map optimization. For the varying
 hyper-parameter $\lambda$, the cross validation results are shown.
 {Alt text: line plot between the rms versus the hyper-parameter for the cross
 validation of the Doppler tomography.}
 }
 \label{f08}
\end{figure}

We fixed the bin size $N=128$. The hyper-parameter $\lambda$ is determined by
cross-validation. The data set is divided into $n=10$ pieces and the optimization
calculation is performed using $n-1$ pieces. The remaining piece is used to validate the
model by evaluating the distance between the data and the model in the energy space. For
each $\lambda$, this is repeated $n$ times for $n$ different selections to derive the
mean and standard deviation of the distribution in the distances
(figure~\ref{f08}). Based on this exercise, we determined $\lambda=0.2$.

\begin{figure}[!hbtp]
 \begin{center}
  \includegraphics[width=1.0\columnwidth,clip]{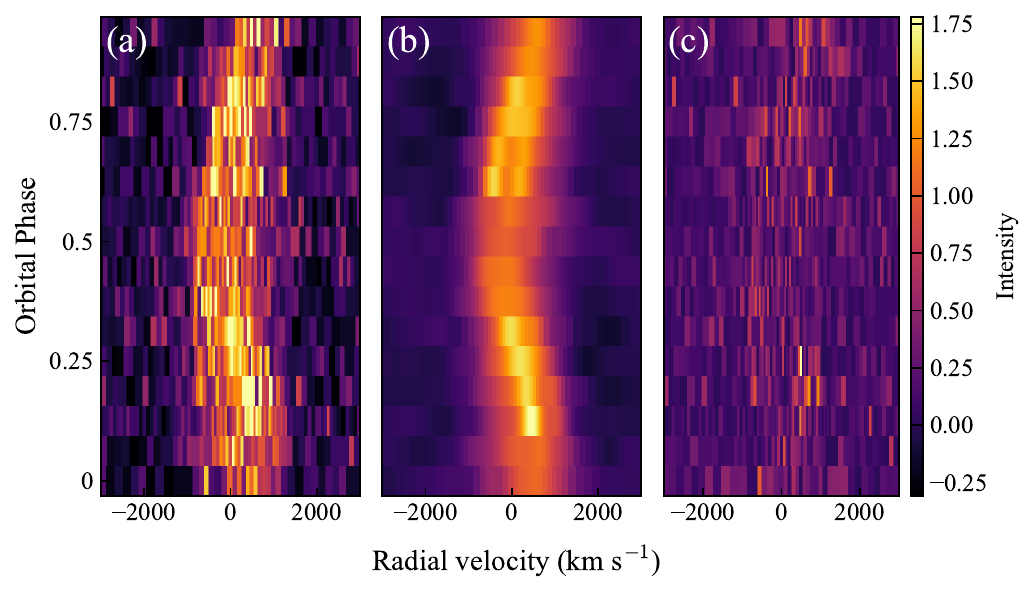}
 \end{center}
 \caption{Trail spectrogram: (a) data, (b) model and (c) residual of the data
 to the model.
 {Alt text: Three maps of the trail spectrogram for the data, model, and residuals.}
 }
 \label{f09}
\end{figure}

\begin{figure}[!hbtp]
 \begin{center}
  \includegraphics[width=1.0\columnwidth,clip]{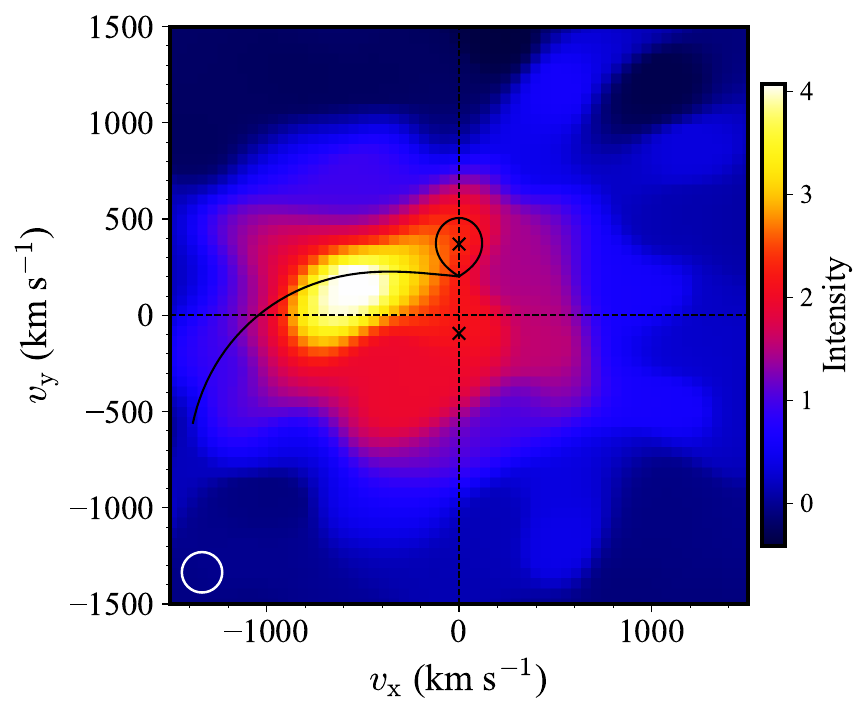}
 \end{center}
 \caption{Velocity map of the Fe~K$\alpha$ line. The position of the NS, the
 companion star, the Roche lobe around it, and the ballistic trajectory from L1 are
 shown. The velocity resolution corresponding to the line spread function is shown at
 the bottom left.
 {Alt text: Velocity map of the Fe~K$\alpha$ line.}
 }
 \label{f03}
\end{figure}

With all these parameters, the resultant solution is visualized as the trail spectrogram
(figure~\ref{f09}) and the velocity map (figure~\ref{f03}). In the former, the residual
(panel c) between the data (panel a) and the model (panel b) is reasonably close to the
null map, though some remaining features remain to be modelled further. In the latter, the
distribution is single-peaked at ($v_{\mathrm{x}}$, $v_{\mathrm{y}}$) $\sim$ (--550,
125)~km~s$^{-1}$ with a dispersion of (432, 253)~km~s$^{-1}$. This is consistent with
the RV curve result for the velocity amplitude and argument. The significant offset from
$\sim$(0, --94)~km~s$^{-1}$ for the NS, an annulus around it for the accretion disk,
$\sim$(0, 370)~km~s$^{-1}$ for the companion star, and the Roche-lobe around it rules
out the possibility that the line is produced in any of these structures. Rather, it is
along the ballistic trajectory from L1 along the accretion stream, suggesting that the
line is produced at a local structure near the stream-disk intersection point.

\section{Discussion}\label{s6}
\begin{figure}[!hbtp]
 \begin{center}
  \includegraphics[width=1.0\columnwidth,clip]{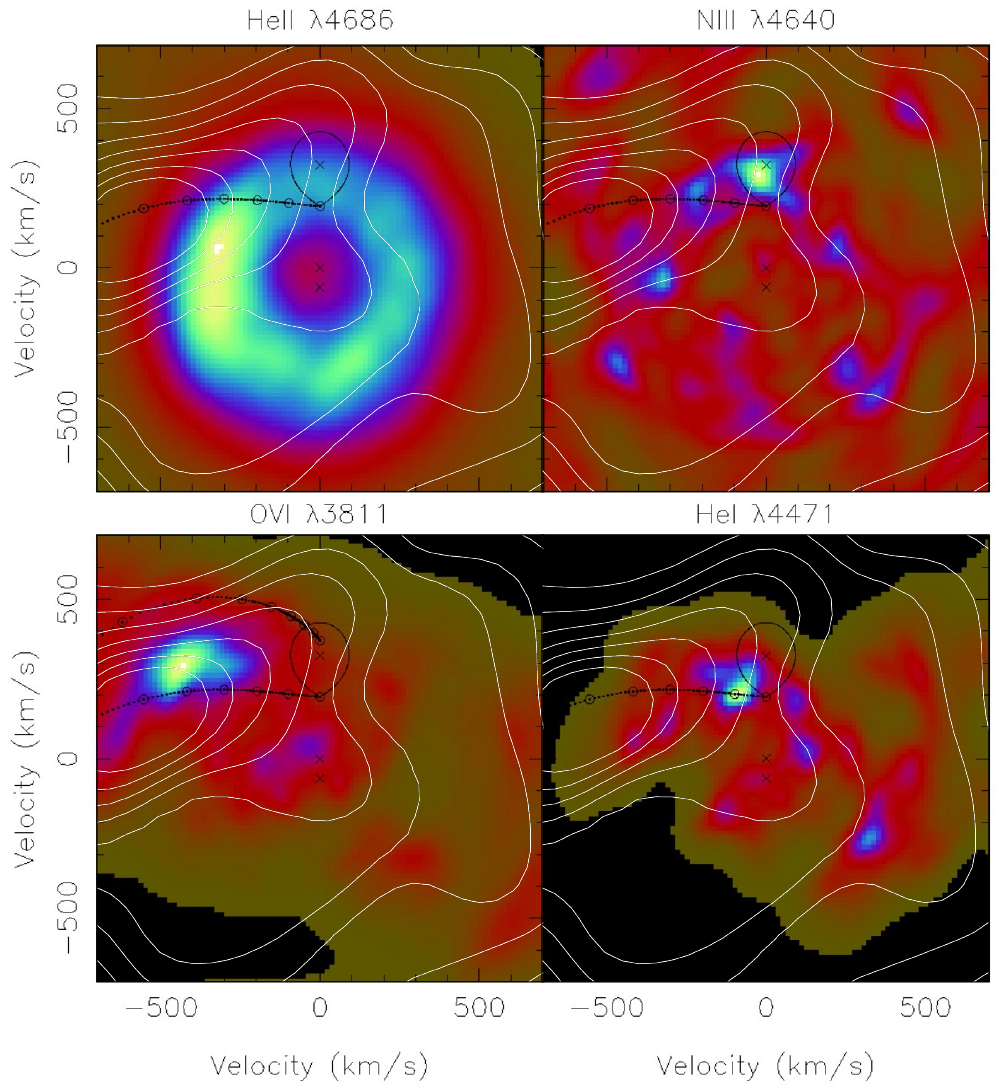}
 \end{center}
 \caption{Comparison of velocity maps. The map of the Fe~K$\alpha$ line (this work) is
 shown with white contours in all four panels. The map of the optical lines
 \citep{casares2003} are shown with color scale in each panel.
 {Alt text: Comparison of velocity maps between optical and X-ray lines.}
 }
 \label{f04}
\end{figure}

To better understand the production site of the Fe K$\alpha$ line, we compared the
velocity map of the Fe K$\alpha$ line (figure~\ref{f03}) with four optical lines that
trace different structures of the accreting system \citep{casares2003} in
figure~\ref{f04}. The overall distribution of the Fe~K$\alpha$ line resembles most that
of the O~VI 3811~\AA{} line (3p $\rightarrow$ 3s) considered to originate from the
stream-disk overflow, and is distinctively different from others. Although no Doppler
tomography was presented, the other O~VI line at 1032/1038~\AA{} (2p $\rightarrow$ 2s
doublet) in the FUV band also shows similar offset in the RV curve
\citep{hutchings2005}. This strongly suggests that the Fe K$\alpha$ line is produced
from the same positional space, as well as the velocity space, with the O VI lines in
this system. The peak of the O~VI map is within 1$\sigma$ dispersion of the Fe K$\alpha$
peak. A possible slight offset between the two peaks is presumably due to different
positions within the overflow.  The fact that both the Fe K$\alpha$ and O VI lines are
eclipsed with an equal fraction with the continuum emission (figure~\ref{f06}b and
Figure 4 in \cite{casares2003}) further supports the idea.

In fact, the co-presence of the Fe~K$\alpha$ line and the O~VI lines is reasonable. The
observed energy of the Fe~K$\beta$ line is $7.074 \pm 0.001$~keV (figure~\ref{f07}),
which is significantly blue-shifted from 7.047~keV expected for Fe~I, while no similar
blue shift is observed for the Fe~K$\alpha$ line. This differential energy shift between
Fe~K$\alpha$ and K$\beta$ is naturally explained for the atomic physical reason that
K$\alpha$ shifts redward while K$\beta$ shifts blueward as the degree of ionization
increases from neutral to Ar-like Fe \citep{nagai2026,palmeri2003a}. The
differential energy is a monotonic function of the ionization degree and the measured
value translates to K-like (Fe~VIII) or Ar-like (Fe~IX). The ionization up to O~VI and
Fe~VIII--IX is likely caused by photons from the NS. 

\begin{figure}[!hbtp]
 \begin{center}
  \includegraphics[width=0.9\columnwidth,clip]{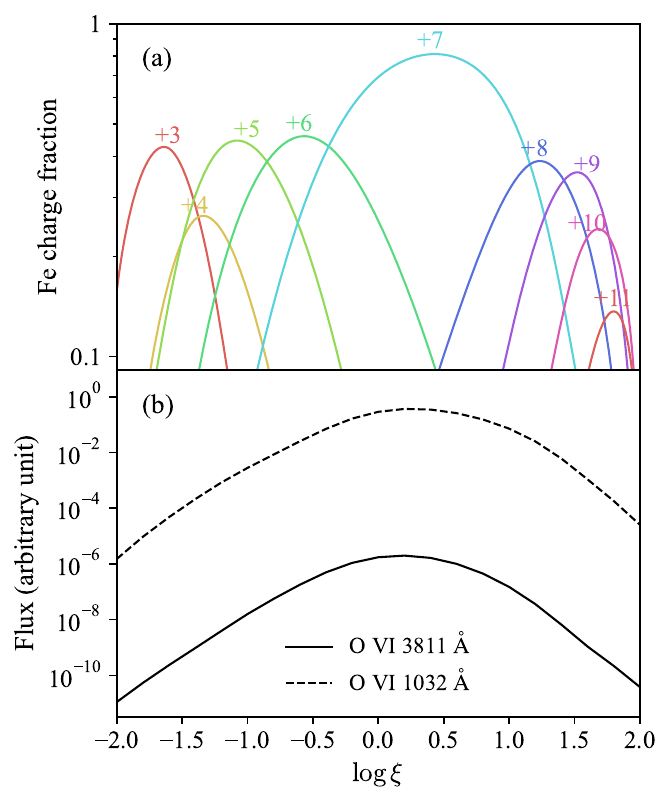}
 \end{center}
 \caption{Ion and line formation as a function of ionization degree. (a)
 Charge fraction of Fe, (b) line intensity of O~VI~3811~\AA{} and (c) O~VI~1032~\AA{} as
 a function of the ionization degree $\xi$ (erg~cm~s$^{-1}$).
 {Alt text: two line plots for the charge state distribution and O VI lines as a
 function of the ionization degree.}
 }
 \label{f10}
\end{figure}

We conducted a radiative transfer calculation using the \texttt{spex} code version
3.08.01 \citep{kaastra1996}\footnote{Available at \url{https://github.com/spex-xray}.}
to examine if the O VI lines and Fe VII--VIII K$\alpha$ line can arise at the same place
in the system. We assume that they are photoionized by the X-ray emission from the NS or
the accretion disk, which is described by a simple model of a broken power-law continuum
attenuated by the interstellar extinction \citep{sasano2014}. We used the intrinsic
shape as an input. The degree of ionization is parameterized by the ionization parameter
$\xi=L_{\mathrm{X}}/nr^{2}$~erg~cm~s$^{-1}$ \citep{tarter1969}, in which
$L_{\mathrm{X}}$ is the X-ray luminosity in 1--1000~Ryd, $n$ is the local density, and
$r$ is the distance from the NS.

We placed a plane-parallel slab with a uniform density $10^{12}$~cm$^{-3}$ and a
thickness of $10^{16}$~cm$^{-2}$. They are intended to be at the low density and the
optically thin limits, but the conclusion is applicable to other settings. As the result
of the radiative transfer calculation, we obtained the charge fraction of Fe in the slab
and the strength of emission lines of interest. We repeated the calculation for varying
$\log{\xi}=$-2 to 2 with a 0.02 dex resolution and obtained the result in
figure~\ref{f10}. At $\log{\xi}=$~0.3, the charge fraction of Fe$^{7+}$ is close to the
maximum, which give rise to Fe VIII K$\alpha$ fluorescence. At the same $\log{\xi}$, the
O VI 3811~\AA{} and 1032~\AA{} lines are at the maximum. This result reinforces the idea
that these lines are formed at the same distance from the NS, hence at the same part of
the accreting system.

\section{Conclusion}\label{s7}
The production site of the Fe K$\alpha$ line has been discussed in many accreting
systems. The broad and narrow lines are distinguished with their equivalent width of
$\sim$1~keV and $\lesssim 100$~eV, respectively. The broad line is presumably from the
inner accretion disk \citep{fabian2005}, while the narrow line, including the one
presented in this paper, is from the outer part of the system.

The production site of the narrow Fe K$\alpha$ line has been attributed to different
structures in different systems, including the outer part of the accretion disk, clumped
dense gas of the disk wind, illuminated surface of the companion star, stellar wind from
the companion star, accretion flow along the magnetic field of NS
\citep{zane2004,Schulz2008,dashwoodbrown2022,kotani1999,kohmura2001}. It may be true
that different systems have different major production sites of the Fe K$\alpha$ line,
but this unconverged view is due in part to the poor quality of the data to date. With
the X-ray microcalorimeter spectrometer and the Doppler tomography technique, we argued
that the Fe K$\alpha$ line in 4U1822 is from the accretion stream-disk overflow. This
approach is generic enough to be applicable to other X-ray lines and in other binary
systems, adding a new and powerful probe to our toolbox to investigate the structures of
these accreting systems.

\begin{ack}
 We appreciate the anonymous reviewer for improving the manuscript. This research made
 use of the JAXA's high-performance computing system JSS3. NS was financially supported
 by the JSPS Core-to-Core Programme (grant number: JPJSCCA20220002) and the Foundation
 for Promotion of Astronomy.
\end{ack}

\bibliographystyle{aa}
\bibliography{ms}{}

\end{document}